\documentstyle[]{article}
\begin{document}
\newcommand{\be}{\begin{equation}}
\newcommand{\ee}{\end{equation}}
\newcommand{\p}{^{\prime}}
\newcommand{\f}{\displaystyle\frac}
\title{Why two formulas for the electric field intensity and the magnetic
induction created by a uniformly moving point charge?}

\author{B.Rothenstein{\footnote{Corresponding author:
bernhard\_rothenstein@yahoo.com}}, I.Zaharie\\ {\small Physics
Department, "Politehnica" University Timi\c soara,}\\ {\small
Pia\c ta Regina Maria, nr.1, 300004 Timi\c soara, Romania}}
\date{}
\maketitle
\begin{abstract}
A derivation of the electric field intensity and of the magnetic
induction generated by a uniformly moving point charge is
presented. The derivation is in accordance with the fact that the
electric and magnetic fields of moving charge are propagating from
the charge with speed $c$ in empty space. The derivation is
tailored for a special audience who knows the transformation of
space, time and fields and Coulomb's law in the rest frame of the
point charge. It avoids the use of Maxwell's equations and the
concept of retarded potential.
\end{abstract}

\section{Introduction}
 \hspace*{6mm}Most university-level introductory physics text-books[1], [2] and
higher level books[3], [4] present the problem of the electric and
magnetic fields generated by an uniformly moving point charge. The
problem can be stated as follows: Consider a point like charge $q$
located at the origin $O'$ of the $K'(X',Y',Z',O')$ reference
frame and at rest relative to it. Reference frame $K'$ moves with
constant speed $\vec {V}$ relative to the $K(X,Y,Z,O)$ reference
frame in the positive direction of the overlapped $OX(O'X')$ axes.
The axes of the two reference frames are parallel to each other.
At the common origin of time in the two frames $(t=t'=0)$ the
origins $O$ and $O'$ of the two frames are shortly located at the
same point in space. Observers from $K'$ detect the presence of a
time independent (static) electric field described by Coulomb's
law. Find out the field generated by the same charge as detected
from the $K$ reference frame.

The problem has a long history that starts with the expressions
for the electric field intensity $\vec E$ and for the magnetic
induction $\vec B$ generated by a point charge moving with
constant velocity $\vec V$ obtained by Oliver Heaviside [5](1888)

\be
\vec{E}=\f{q}{4\pi\varepsilon_0r^3}\f{1-\f{V^2}{c^2}}{\left(1-\f{V^2}{c^2}\sin^2\theta\right)^{3/2}}\vec{r}
\ee

\be \vec{B} = \varepsilon_0 \mu_0 \vec{V}\times \vec{E} \ee where
$\vec{r}$ is the distance between the point where the measurements
of $\vec{E}$ and $\vec{B}$ take place and the charge, representing
the angle between $\vec{r}$ and $\vec{V}$.

Jefimenko [6] reviews the different ways in which (1) and (2) can
be derived. The derivations involve concepts like retarded scalar
and vector potentials (the Lienard-Wiechert potentials) [7], [8],
integrals for retarded potentials of a moving charge
distribution [9], [10], generalized time-dependent Biot-Savart and
Coulomb field laws [11], [12] and Coulomb field laws. All the
mentioned ways to derive (1) and (2) present a high degree of
complexity being hard to teach without mnemonic aids.

Special relativity theory offers a simple derivation of (1) and
(2) [1], [2], [3], [4]. It based on the Lorentz-Einstein transformations
for coordinates, time and fields. From a pedagogical point of view
the teaching of the subject offers the opportunity to stress the
importance of the concept of event and same event.

Resnick [1] offers a solution of the problem we have stated
considering the point $M'(x',y',z')$ of the $K'$ frame where the
point-like charged particle generates the time independent
electric field intensity \be \vec{E}'=
\f{q}{4\pi\varepsilon_0r^3}\f{\vec{r}'}{r'^3}. \ee

Let t' be the time when an observer $R'(x',y',z')$ performs the
measurement being located at the point  $M'(x',y',z')$. Let
$R(x,y,z)$  be an observer of the $K$ frame shortly located in
front of observer $R'$ when both measure the electric field
intensity and the magnetic induction, the first at a time $t$ the
second at a time $t'$. The events associated with the measurements
are $S(x,y,z,t)$ in $K$ and $S'(x',y',z',t')$ in $K'$. Considering
that the space-time coordinates of events $S$ and $S'$ are related
by the Lorentz-Einstein transformations, we ensure the fact that
they refer to the same event. Using the transformations for the
components of the fields we obtain that the electric field
intensity $\vec {E}$ measured by observers from $K$ is given by

\be \vec{E} = D \left[(x-Vt)\vec{1_x}+y\vec{1_y}+z\vec{1_z}\right]
\ee where \be D=
\f{q\gamma}{4\pi\varepsilon_0\left[(x-Vt)^2\gamma^2+y^2+z^2\right]^{3/2}}.
\ee The magnetic induction is given by \be
\vec{B}=\f{\vec{V}\times\vec{E}}{c^2} \ee resulting that observer
$R$ of the $K$ frame detects a time dependent electric field
intensity and magnetic induction as well.

\section{The electric and magnetic fields relative to the
retarded position}

 \hspace*{6mm}  Consider the point charge $q$ located 
at $t' = 0$ at the origin
$O'$. According to the
classical electromagnetic theory the information about the
creation of the field propagates with the same velocity c in all
directions, starting from $O'$ at a time $t'=0$ (Event
$S_0'(0,0,0)$in K'). The event "observer $\displaystyle {R'\left(x'=r' \cos \theta',
y'=r' \sin \theta'\right)}$" detects and measures the electric field
intensity is characterized by the space-time coordinates $S'(x'=r'
\cos \theta', y'=r' \sin \theta', t'=\f{r'}{c})$, $t'=\f{r'}{c}$
representing the time after which the information about the
creation of the fields arrives at the location of $R'$. The
electric field intensity as measured by $R'$ is given by (1). Its
components are

\be E _x' = \f{q}{4\pi\varepsilon_0 r'^2} \cos \theta' \ee

\be E_y' = \f{q}{4\pi\varepsilon_0 r'^2} \sin \theta'. \ee

From the point of view of an observer $R(x=r \cos \theta, y=r \sin
\theta)$ shortly located in front of $R'$ when both measure the
electric field intensity, the same event is characterized by the
space-time coordinates

\be x'=\gamma r \left(\cos \theta - \f{V}{c}\right) \ee \be y' =
y= r \sin \theta \ee

\be r'=\gamma r \left(1- \f{V}{c}\cos \theta \right)
 \ee

\be \cos \theta' = \f{\cos \theta - \f{V}{c}}{1- \f{V}{c}\cos
\theta}\ee

\be \sin \theta' = \f{\gamma \sin \theta}{1- \f{V}{c}\cos \theta}
\ee

\be t' = t \gamma \left(1- \f{V}{c}\cos \theta \right). \ee

The components of the electric field intensity measured by
observer $R$ are

\be E_x = E_x'= \f{q\left(1-\f{V^2}{c^2}\right)}{4 \pi
\varepsilon_0r^2} \f{\cos \theta -\f{V}{c}}{\left(1-\f{V}{c}\cos
\theta \right)^3} \ee

\be  E_y = \gamma E_y'= \f{q\left(1-\f{V^2}{c^2}\right)}{4 \pi
\varepsilon_0r^2} \f{\sin \theta}{\left(1-\f{V}{c}\cos \theta
\right)^3}   . \ee

The electric field measured by $K$ is \be \vec{E} =
E_x\vec{1_x}+E_y \vec{1_y}= \f{q\left(1-\f{V^2}{c^2}\right)}{4 \pi
\varepsilon_0r^3}\f{\vec{r}-\f{r
\vec{V}}{c}}{\left(1-\f{\vec{V}\vec{r}}{rc}\right)^3}    .\ee

The magnetic induction detected by $R$ is

\be \vec{B}=\varepsilon_0 \mu_0 \vec{V}\times\vec{E}  . \ee
\hspace*{6mm} Rosser [13] derives (16) and (17) establishing a
relationship between the lengths of the position vectors that
define the position of the point where the measurements are
performed from the "retarded position" and the "present position"
of the field producing charge and using the transformation of
fields.

Jefimenko [6] derives (17) and (18) based exclusively on general
electromagnetic field equations without making use of
transformation equations. The derivation is not very simple and
not transparent enough making its teaching hard without using
mnemonic aids.
\section{Conclusions}
The derivation of the electric field intensity and of the magnetic
induction of a point charge detected from a reference frame
relative to which it moves with constant velocity we have
presented above is based on the knowledge of transformation
equations for coordinate, time and fields. As compared with other
derivations which use or do not use special relativity our
derivation can be considered as a "two line" one.

\end{document}